\documentstyle[preprint,aps]{revtex}
\date{February 5, 1995}

\def\s{\sigma}
\def\Tr{{\rm Tr}}

\def\dm{\delta M}
\def\se{\sigma_\omega}
\def\bPhi{{\bar \Phi}}
\def\no{\noindent}
\def\a{\alpha}

\begin{document}
\draft
\title{\bf Two-Dimensional Electrons in a Strong Magnetic Field with
Disorder: Divergence of the Localization Length}
\author{K. Ziegler}
\address{Institut f\"ur Theorie der Kondensierten Materie,
Universit\"at Karlsruhe, Physikhochhaus, D-76128 Karlsruhe, Germany}
\maketitle
\begin{abstract}
Electrons on a square lattice with half a flux quantum per plaquette are
considered. An effective description for the current loops is given by
a two-dimensional Dirac theory with random mass. It is shown that
the conductivity and the localization length can be calculated from a product
of Dirac Green's functions with the {\it same} frequency. This implies
that the delocalization of electrons in a magnetic field is due to a
critical point in a phase with a spontaneously broken discrete symmetry.
The estimation of the localization length is performed for a generalized
model with $N$ fermion levels using a $1/N$--expansion and the Schwarz
inequality. An argument for the existence of two Hall transition points
is given in terms of percolation theory.
\end{abstract}
Lattice models for the two-dimensional electron gas in a strong magnetic
field are of increasing interest because of recent experiments with GaAs
heterostructures \cite{sha}. Moreover, lattice models offer new ways
to the study of the quantum Hall transition. The purpose of this article
is to discuss a new concept for the investigation of the localization length
for the lattice model.

The tight-binding Hamiltonian for non-interacting electrons on a square
lattice with magnetic flux $\phi$ reads in Landau gauge
$$H=-t\sum_r[e^{2i\pi y\phi/\phi_0}c^+(r+e_x)c(r)+c^+(r+e_y)c(r)+h.c.]+
\sum_rV(r)c^+(r)c(r).\eqno (1)$$
$e_{x,y}$ are lattice unit vectors, and
$c^+$ and $c$ are fermion creation and annihilation operators, respectively.
$V(r)$ is a potential which represents an additional structure or disorder
on the lattice.
The dispersion for $V=0$ and half a flux quantum per plaquette ($\phi=
\phi_0/2$) is $E=\pm2t\sqrt{\cos^2k_1+\cos^2k_2}$. A linear (relativistic)
approximation for $E\sim0$ around the four nodes $k_j=\pm\pi/2$ is possible.
For $H$ this is still very complicated \cite{fi}, and a further simplification
is
useful to lift the degeneracy of the four nodes. This can be achieved by the
introduction of a next nearest neighbor hopping term and a staggered
potential in $H$ \cite{lud,osh}. This modification is motivated by the network
model \cite{chalk}.
Expansion of $k=(\pi/2,\pi/2)+p$ for small $p$ vectors leads to the large
scale approximation by the Dirac Hamiltonian $H'=\sigma\cdot p+\sigma_3M$ with
Pauli matrices $\s_j$. Disorder, originally introduced in $H$ by a staggered
random potential $V$, is described in $H'$ by a random mass $M$ \cite{lud}.
Physical quantities can be obtained from the Green's function
$$G(\omega)\equiv\pmatrix{
G_{11}&G_{12}\cr
G_{21}&G_{22}\cr
}=\pmatrix{
i\omega+M&h\cr
h^+&i\omega-M\cr
}^{-1},\eqno (2)$$
where the Fourier component of $h$ is $p_1-ip_2$. For instance, the
frequency-dependent conductivity is given by Kubo's formula
$$\sigma_{xx}(\omega)={e^2\over h}\omega^2\sum_rr^2\langle G_{jj'}(r,0;i\omega)
G_{j'j}(0,r;-i\omega)\rangle,\eqno (3)$$
where $G_{jj'}(r,0;i\omega)G_{j'j}(0,r;-i\omega)=|G_{jj'}(r,0;i\omega)|^2$.
The localization length $\xi_l$ is defined as the decay of the function
$C_{jj'}(r,\omega)\equiv\langle|G_{jj'}(r,0;i\omega)|^2\rangle$ in space. There
exists a relation of the above expression,
which is composed of Green's functions at frequencies with {\it opposite sign}
(retarded and advanced Green's functions), with a product of Green's functions
at the {\it same} frequency. This follows from the block matrices in (2)
$$G_{jj}(r,r';-i\omega)=-G_{j'j'}(r',r;i\omega),\ \ \
G_{jj'}(r,r';-i\omega)=-G_{jj'}(r',r;i\omega)\ \ \ (j\ne j').\eqno (4)$$
This identity reflects the Lorentz-covariance of the Dirac theory. It implies
$|G_{jj}(r,r';i\omega)|^2$ $=-G_{jj}(r,r';i\omega)G_{j'j'}(r,r';i\omega)$
and
$|G_{jj'}(r,r';i\omega)|^2=-G_{j'j}(r,r';i\omega)G_{jj'}(r,r';i\omega)$.
That means only the Green's functions with one frequency is required
for the evaluation of transport or localization properties in the present
model.

The physics of the electrons can be understood as the statistics of current
loops, created by the magnetic field. Depending on the potential there are
local current loops with two different directions. This observation is
central for the understanding of the Hall transition discussed
subsequently. The current loops are composed from the Dirac fermions by the
creation and annihilation of particle-hole pairs.  The direction can be
reversed globally by a time-reversal transformation. In terms of the Green's
function this means $G(\omega)\to-\s_3G(\omega)\s_3$. In general the system is
not invariant under this transformation because one direction of the currents
is favored. The favored direction characterizes the electronic state of a Hall
plateau. The transition between the Hall plateaux corresponds to a symmetric
point with $\omega=M=0$ which is also a critical point
with a divergent localization length. The discrete symmetry at this critical
point plays a fundamental role for the lattice electrons in a magnetic field.
Taking now a random mass, which is symmetrically distributed around zero, this
symmetry is spontaneously broken for the average Green's function \cite{zie1}.
It will
be discussed in the following that the localization length also diverges at
two critical points in the presence of disorder where the symmetry is broken.
This is an extension of a previous work \cite{zie2} where the Hall conductivity
was calculated. Critical points with a divergent localization
length in a symmetry
broken regime are known from Anderson localization in systems without
magnetic field and dimensionality $d>2$. However, the physics of the latter is
different because there are no current loops. As a manifestation of this
difference a {\it continuous} symmetry appears \cite{weg} instead of discrete
symmetry
found for localization in the presence of a magnetic field. The
large scale approximation for the continuous symmetry leads to a non-linear
sigma model contrary to the Dirac theory in the presence of the
magnetic field.

The Schwarz inequality can be applied to get a lower bound for $C$:
$$
|\langle G_{jj''}(r,0;i\omega)G_{j'j'''}(0,r;i\omega)\rangle|^2\le
\langle|G_{jj''}(r,0;i\omega)|^2\rangle\langle|G_{j'j'''}(0,r;i\omega)|^2
\rangle \eqno (5)$$
with $j''=j$, $j'''=j'$ or $j''=j'$, $j'''=j$. Writing $C'_{jj}(r,\omega)
\equiv\langle G_{jj}(r,0;i\omega)G_{j'j'}(0,r,i\omega)
\rangle$ and $C'_{jj'}(r,\omega)\equiv\langle G_{jj'}(r,0;i\omega)G_{j'j}
(0,r,i\omega)\rangle$ one has $|C_{jj'}'(r,\omega)|\le|C_{jj'}(r,\omega)|$.
$C'$ will be used subsequently because it easier to calculate than $C$.
The average correlation functions are translational invariant. Therefore, the
corresponding Fourier components ${\tilde C}(k,\omega)$ can be used to
calculate the localization length
$$\xi_l\propto\sqrt{-{\nabla_k^2{\tilde C}(k,\omega)\over{\tilde C}
(k,\omega)}}\Big|_{k=\omega=0}.
\eqno (6)$$
A simple calculation for the pure system gives $\xi_l=|M|^{-1}$. To evaluate
the correlation function $C'(r,\omega)$ a generalization of
the Hamiltonian $H'$ is introduced which describes $N$ levels of fermions per
site \cite{zie2}:
$H^{\alpha \alpha'}=H_0^{\alpha \alpha'}-\dm_{r}^{\alpha \alpha'}
\s_3$ ($\alpha ,\alpha'=1,2,...,N$)
with $H_0^{\alpha \alpha'}=(\s\cdot p+\langle M\rangle\s_3)\delta^{\alpha
\alpha'}$. The distribution of the real symmetric
random matrix (a Gaussian Orthogonal Ensemble \cite{mehta}) $\dm$ is given by
$\langle\dm_{r}^{\alpha \alpha'}\dm_{r'}^{\alpha'' \alpha'''}\rangle
=(g/N)\delta^{\alpha \alpha'''}\delta^{\alpha' \alpha''}\delta_{r,r'}$.
That means only random fluctuations couple the $N$ different Dirac systems.
The product of two Green's functions with the same frequency can be expressed
as a functional integral \cite{neg}
$$G_{j,j''}^{\a\a'}(r,r';i\omega)G_{j',j'''}^{\a''\a'''}(r'',r''';i\omega)
=-\int{\bar \Psi}_{r',j''}^{\a'}\Psi_{r,j}^{\a}
{\bar\chi}_{r''',j'''}^{\a'''}\chi_{r'',j'}^{\a''}
\exp(-S_1)\prod_r d\Phi_r d{\bar \Phi_r}\eqno (7)$$
with the action (sum convention for the level index $\a$)
$$S_1=i\se [-(\Phi,(H_0+i\omega\s_0)\bPhi)+\sum_r\dm_{r}^{\a\a'}
(\Phi_{r}^{\a'}\cdot\s_3\bPhi_{r}^{\a})].\eqno (8)$$
$\se ={\rm sign}(\omega)$ and the field is
$\Phi_{r,j}^\a=(\Psi_{r,j}^\a,\chi_{r,j}^\a)$.
The first component is Grassmann and the second is complex.
The complex component is added to normalize the functional integral in (7).
It also provides a transparent representation of the product of Green's
functions with the same frequency as required for $C$ and $C'$.
Averaging with Gaussian distributed fluctuations yields $\langle\exp(-S_1)
\rangle=\exp(-S_2)$ with
$$S_2=-i\se(\Phi,(H_0+i\omega\s_0)\bPhi)+ {g\over N}\sum_r(\Phi_{r}^\a\cdot
\s_3 \bPhi_{r}^{\a'})(\Phi_{r}^{\a'}\cdot\s_3 \bPhi_{r}^\a).\eqno (9)$$
This action can be mapped onto another effective field theory \cite{zie3}:
The
interaction in $S_2$ can also be created by complex $2\times2$-matrix fields
$Q$, $P$ and a complex Grassmann field $\psi$ which couple to the
composite field $\sum_{\a=1}^N\Phi_{r}^{\a}\bPhi_{r}^{\a}$. The new field
does not depend on $\a$, the level index of the $N$ level fermions.
The level degree of freedom can be eliminated by integrating out the field
$\Phi$ in the functional integral. This leads to $\exp(-NS(Q,P,\psi))$
with the action \cite{zie3}
$$S(Q,P,\psi)={1\over g}\sum_r[\Tr_2(Q_r\s_3)^2+\Tr_2(P_r\s_3)^2]
+\log\det(H_0-2Q+i\omega\s_0)$$
$$-\log\det(H_0+2iP+i\omega\s_0)+{1\over N}\sum_{\mu,\mu'=1}^4
\int ({\bf I})_{\mu,\mu'}(k)\psi_{k,\mu}{\bar\psi}_{-k,\mu'}d^2k+O(N^{-2}).
\eqno (10)$$
Since the number of fermion levels $N$ appears in front of the new action,
the limit $N\to\infty$ corresponds to a saddle point integration for the fields
$Q$ and $P$.
${\bf I}(k)$ is the matrix of the Gaussian fluctuations around the
saddle point solution $Q=Q_0+\delta Q$ and $P=iQ_0+\delta P$, where $Q_0$ is
the $N\to\infty$ solution $Q_0=-(1/2)[i\eta\s_0+M_s\s_3]$. $M_s$ is a shift
of the average Dirac mass and $\eta$ shifts the frequency in the Green's
function. Introducing $m=\langle M\rangle+M_s$ as the effective (renormalized)
Dirac mass the imaginary shift is $\eta=\s_\omega\sqrt{
e^{-2\pi/g}-m^2}$ for $|m|\le m_c\equiv e^{-\pi/g}$ and zero otherwise
\cite{zie3}.

The details of the derivation of $S(Q,P,\psi)$ from $S_2$ and the evaluation
of saddle point integration can be found in Ref. \cite{zie3}. However, (10)
can also be reconstructed using the fact that $\int\exp(-S_1)d...=1$.
The $N\to\infty$-limit gives $Q_0=-iP_0$ which implies that the large $N$-terms
cancel each other in the action (10). First order fluctuations do not
contribute at the saddle point. And second order fluctuations
appear as quadratic forms of $\delta Q$ and $\delta P$ with the matrix
${\bf I}$. Therefore, the condition $\int\exp(-NS(Q,P,\psi))d...=1$
is satisfied if a quadratic form of ${\bf I}$ appears with a complex
Grassmann field. This result reflects the supersymmetric construction of the
functional integral.

For the asymptotic behavior of a large correlation length it is sufficient
to consider $k\sim0$. If $k=0$ the matrix ${\bf I}$ reads
$${\bf I}(k=0)=\pmatrix{
1/g-2\a\mu^2&0&0&2\beta\cr
0&2/g+4\a|\mu|^2&0&0\cr
0&0&2/g+4\a|\mu|^2&0\cr
2\beta&0&0&1/g-2\a\mu^{*2}\cr
},\eqno (11)$$
where $\mu=m+i\eta$, $\a=\int (|\mu|^2+k^2)^{-2}d^2k/4
\pi^2$ and $\beta=\int k^2(|\mu|^2+k^2)^{-2}d^2k/4\pi^2$.
The inner $2\times2$ matrix of ${\bf I}(k=0)$ is diagonal with
positive matrix elements $2/g+4\a|\mu|^2$. It leads always to a finite
correlation length. However, this part of ${\bf I}$ does not contribute to
$C'$ because $C'_{jj'}={\bf I}^{-1}_{j+2(j-1),j'+2(j'-1)}$ \cite{zie3}.
Therefore,
only the projected submatrix ($P_{14}$ is the projector on the indices $1,4$)
$$P_{14}{\bf I}^{-1}P_{14}
=\pmatrix{
1/g-2\mu^2\a+\mu^2A_1{k}^2 &2\beta+B_1{k}^2\cr
2\beta+B_1{k}^2&1/g-2\mu^{*2}\a+\mu^{*2}A_1{k}^2 \cr
}^{-1}\eqno (12)$$
with positive constants $A_1$, $B_1$
is of interest here. It describes the critical behavior of the model
due to one vanishing eigenvalue for $k=0$ at two critical points $m=\pm m_c$.
The critical points are also characterized by a vanishing
imaginary part of the saddle point $Q_0$.

The Fourier components of $C'$ are of the form ${\tilde C}'(k,\omega=0)=
(2\beta+bk^2)/(a'+b'k^2)$, where $b$ and $b'$ are finite constants and
$a'=|1/g-2\a\mu^2|^2-4\beta^2$.
Equ. (6) yields for the correlation length $\xi$ of $C'$
$$\xi\propto2(b'/a'-b/2\beta)^{1/2}\sim2\sqrt{b'/a'}.\eqno (13)$$
Because $b'$ is always finite the divergence of the correlation
length is determined by $1/a'$. The behavior of $a'$ distinguishes the regime
of $|m|\le e^{-\pi/g}$ and $|m|>e^{-\pi/g}$. This leads to
$$
\xi\sim\xi_0 (1/g+{1\over\pi}\log|m|)^{-1/2}\ \ \ {\rm for}\ \ |m|>e^{-\pi/g}
\eqno (14a)$$
and
$$
\xi\sim\xi_0'(e^{-2\pi/g}-m^2)^{-1/2}\ \ \ {\rm for}\ \ |m|\le e^{-\pi/g}.
\eqno (14b)$$
Thus $\xi(m)$ diverges linearly at $m=\pm e^{-\pi/g}$ (then $\eta=0$
according to the saddle point solution), and it is finite elsewhere.
The existence of two critical points, instead of the single critical
point in the perturbative approach of Ref.\cite{lud}
is a consequence of the spontaneous symmetry breaking due to disorder
which creates a non-vanishing density of states between the particle
and hole band of the Dirac theory. This non-perturbative phenomenon
was discussed in Ref.\cite{zie1,zie2,zie3}.

The correlation function $C$ cannot be calculated directly from
$S(Q,P,\psi)$ because it not expressable as a simple form in the
large $N$ limit. Therefore, only $C'$ can be used as a lower bound for the
decay of $C$ due to inequality (5). As a consequence, $\xi$ bounds
the localization length $\xi_l$ from below.
An alternative approach, based on a different
effective field theory with $N=1$, indicates also a square root divergent
localization length \cite{zie4}. The exponent $\nu\approx1$ for the
localization
length was found experimentally for the metal-insulator transitions
(i.e. near the mobility edge) in AlGaAs/GaAs heterostructures \cite{sha}
and in Si MOSFETs \cite{dol}.

On the other hand, numerical results for network models and lowest Landau
level approaches
\cite{chalk,huck,mie,huo} suggest that the localization length
$\xi_l$ diverges at the critical point with an exponent $\nu\approx7/3$.

\no
The existence of
two transition points is in disagreement with the single transition point of
other approaches to the integer Hall transition
\cite{chalk,huck,mie,huo,pruisk}.
This is either due to different models (Landau level approach
versus lattice fermions) or due to the fact that the two transition points
were not seen because of limitations of the numerical resolution.
(The distance of the two transitions is $2e^{-\pi/g}$ which is extremely small
for weak disorder. It can be hidden by finite size effects.)
Two transition points are also supported by the percolation picture for the
lattice model. This can be seen considering a lattice where the
space-dependent Dirac mass $M_r$ is either zero with
probability $p$ or $M$ with probability $1-p$.
The correlation length $\xi$ is
infinite for $p=1$ because $M=0$ is the symmetric point.
For $p<1$ not all lattice sites have a zero mass.
However, there is typically an infinite cluster of lattice
sites with zero mass for $p_c<p<1$ according to percolation theory
\cite{essam}.
The correlation length of the classical percolation system is finite for
$p<p_c$ and diverges like $\xi\sim\xi_0
(p_c-p)^\nu$ with the critical exponent $\nu=4/3$.
The mean value of $M_r$ at the critical point
$p_c$ is $\langle M\rangle=(1-p_c)M
\equiv M_c$. It seems plausible that its value is
non-zero even if a renormalization ($\langle M\rangle\to m$) by quantum
fluctuations of the fermions is taken into account. Application of the
transformation $G\to\s_3G\s_3$ to the Green's function $G(\omega=0)$ replaces
$M$ by $-M$. Thus $-M_c$ is also a critical point.

In conclusion, it was shown that the transport of electrons on a square
lattice in a strong magnetic field and random potential can be described by a
product of Dirac Green's function with the same frequency. Therefore, the
physics is characterized by a discrete symmetry. The localization length
diverges at two different critical points.
The critical exponent was estimated by $\nu\ge1/2$.

\end{document}